\documentclass[conference]{IEEEtran}
\IEEEoverridecommandlockouts
\usepackage{cite}
\usepackage{amsmath,amssymb,amsfonts}
\usepackage{algorithmic}
\usepackage{graphicx}
\usepackage{textcomp}
\usepackage{xcolor}

\usepackage{algorithmic}
\usepackage{algorithm}
\usepackage[caption=false,font=normalsize,labelfont=sf,textfont=sf]{subfig}

\def\BibTeX{{\rm B\kern-.05em{\sc i\kern-.025em b}\kern-.08em
    T\kern-.1667em\lower.7ex\hbox{E}\kern-.125emX}}
\begin{document}

\title{Second-Order Analysis of CSMA Protocols for Age-of-Information Minimization\thanks{This material is based upon work supported in part by NSF under Award Numbers ECCS-2127721 and CCF-2332800, the U.S. Army Research Office under Grant Number W911NF-22-1-015, and the U.S. Office of Naval Research under Grant Number N000142412615. Portions of this research were conducted with the advanced computing resources provided by Texas A\&M High Performance Research Computing.}}

\author{\IEEEauthorblockN{Siqi Fan}
\IEEEauthorblockA{\textit{Dept. of Electrical \& Comp Engr} \\
\textit{Texas A\&M University}\\
College Station, USA \\
sf26372@tamu.edu}
\and
\IEEEauthorblockN{I-Hong Hou}
\IEEEauthorblockA{\textit{Dept. of Electrical \& Comp Engr} \\
\textit{Texas A\&M University}\\
College Station, USA \\
ihou@tamu.edu}
}

\maketitle

\begin{abstract}
This paper introduces a general framework to analyze and optimize age-of-information (AoI) in CSMA protocols for distributed uplink transmissions. The proposed framework combines two theoretical approaches. First, it employs second-order analysis that characterizes all random processes by their respective means and temporal variances and approximates AoI as a function of the mean and temporal variance of the packet delivery process. Second, it employs mean-field approximation to derive the mean and temporal variance of the packet delivery process for one node in the presence of interference from others. To demonstrate the utility of this framework, this paper applies it to the age-threshold ALOHA policy and identifies parameter settings that outperform those previously suggested as optimal in the original work that introduced this policy. Simulation results demonstrate that our framework provides precise AoI approximations and achieves significantly better performance, even in networks with a small number of users.
\end{abstract}



\section{Introduction}

The growing demand for real-time communication in distributed systems, such as Internet of Things (IoT) networks, wireless sensor networks, and other time-sensitive applications, has increased the focus on optimizing data freshness. While traditional metrics such as throughput and reliability remain important, they often fail to capture the need for timely updates of information. The Age-of-Information (AoI) metric addresses this challenge by measuring the freshness of data at the receiver, quantifying the time elapsed since the most recent update was received \cite{kaul2011minimizing}. As a result, AoI has become a key performance indicator for networks where timely data delivery is critical.

Optimizing AoI in random access networks that use slotted Carrier Sense Multiple Access (CSMA) protocols presents significant challenges. These networks require multiple users to share a communication channel, where each transmission attempt is only successful if no other user transmits in the same time slot. Each user’s transmission strategy is governed by a Markov process, with state transitions depending on the results of channel sensing or the reception of an acknowledgment (ACK).

In slotted CSMA networks, the use of sensing mechanisms and ACKs introduces dependencies between users' transmission decisions, as each user’s decision is influenced by the decisions of other uses in the previous slot. These dependencies complicate the analysis because it becomes necessary to track the state of each user, rendering traditional analytical methods intractable as the number of users increases. This complexity contrasts with previous studies, such as those by Yates \textit{et al.} \cite{yates2017status} and Talak \textit{et al.} \cite{talak2018distributed}, which assume independent user actions and focus on simpler protocols like slotted ALOHA.

To date, much of the existing research has focused on specific strategies for AoI optimization in random access networks, such as age-threshold policies or other age-aware transmission strategies. For instance, Chen \textit{et al.} \cite{chen2020age} proposed an age-sensitive random access protocol where nodes transmit only when their instantaneous AoI exceeds a predefined threshold, an approach later extended to CSMA networks by Yavascan \textit{et al.} \cite{yavascan2021analysis}. Similarly, Chen \textit{et al.} \cite{chen2022age} introduced a distributed transmission strategy based on an age-gain metric, quantifying the reduction in AoI following successful packet delivery. Further approaches include the index-prioritized random access scheme by Sun \textit{et al.} \cite{sun2019closed} and the minislotted threshold ALOHA protocol analyzed by Ahmetoglu \textit{et al.} \cite{ahmetoglu2022mista}, which is designed for high payload scenarios. Moradian \textit{et al.} \cite{moradian2024age} considered dynamic frame slotted ALOHA with an age-gain threshold. Kadota \textit{et al.} \cite{kadota2021age} developed a framework to optimize average AoI in networks with stochastic packet generation, while Wang \textit{et al.} \cite{wang2024analytical} employed stochastic hybrid systems (SHS) to analyze AoI in CSMA networks.

Despite these advances, a unified framework that accommodates arbitrary Markov process-based transmission strategies for optimizing AoI in slotted CSMA remains a challenge. Many existing works are limited to special cases, such as when the number of users approaches infinity. A more general approach is needed to extend the analysis and optimization of AoI to a wider variety of CSMA protocols and network sizes.

To address these challenges, we propose a general framework that combines second-order analysis and mean-field approximation to model and optimize AoI in random access networks using CSMA protocols. The second-order analysis approximates AoI by considering both the mean and temporal variance of packet delivery processes, transforming the problem of AoI optimization into one of optimizing these two metrics. The mean-field approximation models user interference, enabling the derivation of the mean and variance of the packet delivery process for each user while accounting for dependencies introduced by sensing and acknowledgment mechanisms. This combined approach allows for flexible analysis and optimization across various CSMA protocols, including those with age-threshold policies and other Markov-based strategies.

Our primary contribution is the development of the scalable and flexible framework for analyzing and optimizing Age-of-Information (AoI) in random access networks under arbitrary CSMA protocols. This framework offers improved flexibility and performance in AoI minimization. We apply the framework specifically to the age-threshold policy and demonstrate its accuracy in scenarios with small numbers of users, extending beyond the limitations of asymptotic analyses in previous works. Simulation results confirm that the parameters derived from our approach consistently outperform those proposed in \cite{yavascan2021analysis}, particularly in practical settings with fewer users, underscoring the effectiveness and robustness of our method.

The remainder of this paper is structured as follows. Section~\ref{sec:system} introduces the system model and problem formulation. In Section~\ref{sec:solution}, we detail our framework, which combines second-order analysis and mean-field approximation to optimize AoI. Section~\ref{sec:results} presents the practical results of applying the proposed approach to parameter settings in recently proposed policies, showcasing its effectiveness. Finally, Section~\ref{sec:conclusion} concludes the paper.


\section{System Model}
\label{sec:system}

\begin{figure}[htbp]
\centerline{\includegraphics[scale=0.3]{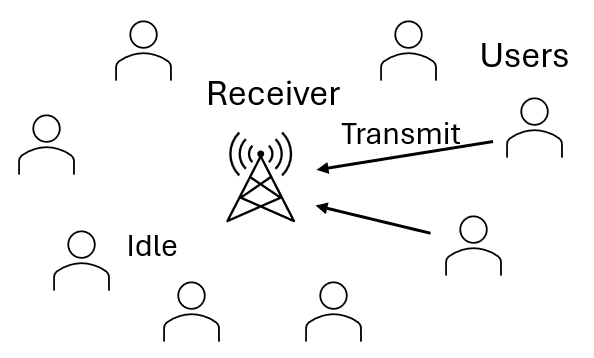}}
\caption{Network Model Example.}
\label{fig:system}
\end{figure}

We analyze a random access network consisting of $N$ symmetric users and a single receiver. The goal is to analyze how users manage their transmissions and how network performance can be optimized when users interfere with each other. The network operates in slotted time, where each time slot provides an opportunity for users to transmit their packets and each packet transmission is completed within a single time slot. If only one user transmits during a slot, the transmission is successful; otherwise, a collision occurs, resulting in no data received by the receiver. Figure~\ref{fig:system} illustrates the network setup.

\vspace{-5pt}

In this setup, users follow a slotted CSMA protocol, with their behavior modeled as a Markov process that consists of two kinds of states: a transmission state and several idle states, whose behaviors are described below:
\begin{itemize}
    \item Idle states: When a user is in one of the idle states, it does not make any transmission. The user senses the channel to see if there are any other transmissions going on. Based on the sensing result, the user transition to another state.
    \item Transmission state: When a user is in the transmission state, it transmits a packet containing its latest information. The user then waits for an ACK. Based on whether it receives an ACK or not, the user transitions to a different state.
\end{itemize}

For simplicity, we assume each user can be in one of $S$ states, where the first state is the transmission state, and the others are idle states. The transition dynamics between these states are crucial for balancing network load and minimizing collisions.

The users transition between these states based on the sensing results, which influence their Markov process transitions. A user’s decision to transmit or remain idle depends on whether the channel is sensed as idle or busy, and this decision is captured by the parameter $\alpha$, defined as follows:
\begin{itemize}
    \item $\alpha = 0$: The channel is sensed as idle (i.e., no other user transmitted) or the user received an ACK from the previous time slot. In this case, the user follows the Markov transition matrix $M_0$, which may increase the likelihood of transitioning to a transmission state.
    \item $\alpha = 1$: Another user transmitted in the previous time slot, leading to a collision if the user transmitted. The user then follows the transition matrix $M_1$, which likely keeps the user in an idle or backoff state to avoid collisions.
\end{itemize}

This setup models the user’s state evolution as a function of the sensing results and internal protocol dynamics, with multiple idle and transmission states.

Each user’s performance is evaluated by its Age of Information (AoI). Every user generates a new packet in each time slot, and the AoI of user $n$ at time $t$ is recursively defined as:
\begin{align*}
    AoI_{n,t} := \begin{cases}
                    1, & \text{if users $n$ successfully delivers} \\
                    & \text{a packet at time $t$,} \\
                    AoI_{n,t-1} + 1, & \text{otherwise}.
                \end{cases}
\end{align*}

\begin{figure}[htbp]
\centerline{\includegraphics[scale=0.29]{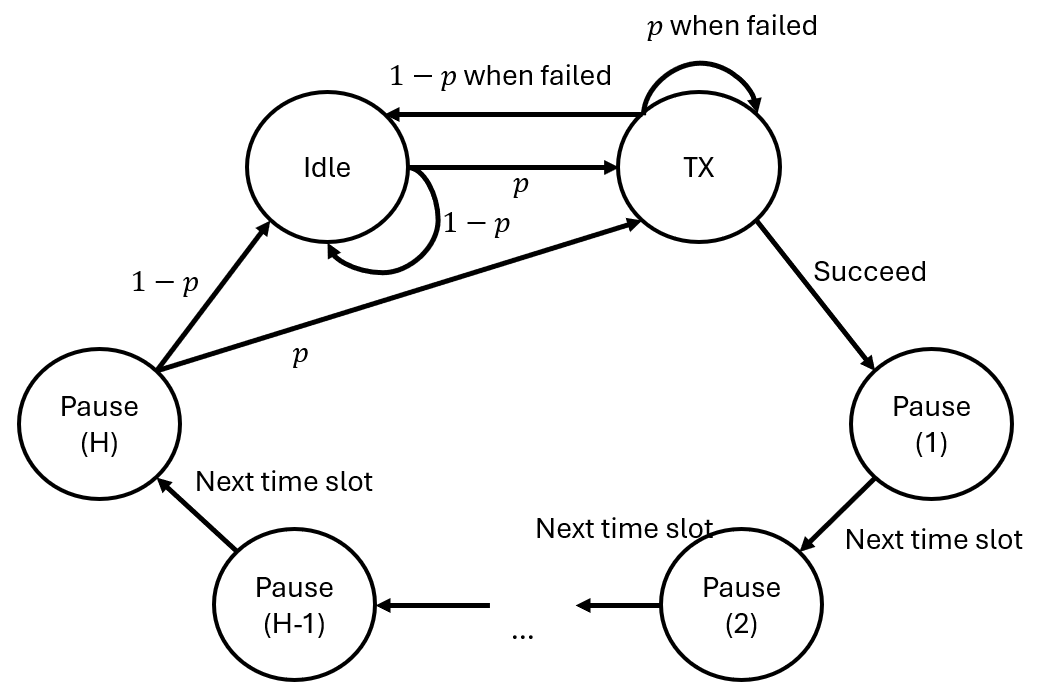}}
\caption{Markov Process for the TA Policy.}
\label{fig:ata}
\end{figure}

To demonstrate this concept, consider a specific strategy known as the age-threshold ALOHA policy, introduced in recent literature \cite{yavascan2021analysis} to minimize AoI. In this strategy, each user pauses its transmissions until the AoI of the user exceeds a predefined threshold $H$. Once the AoI surpasses $H$, the user begins transmitting with a constant probability $p$ in each slot until a successful transmission occurs. After the success, the user pauses transmissions again until the AoI exceeds the threshold $H$. This process can be described as a Markov process, as shown in Figure.~\ref{fig:ata}.

The transition matrices $M_0$ and $M_1$ for this process are of size $(H+2) \times (H+2)$ and are defined as follows, where state 1 is the transmission state (TX), state 2 is the idle state for ALOHA, and states 3 to $H+2$ are pausing states with different ages.

\begin{align*}
    M_0 = \begin{bmatrix}
            p & 1-p & 0 & 0 & 0 & ... & 0 & 0\\
            p & 1-p & 0 & 0 & 0 & ... & 0 & 0\\
            0 &  0  & 0 & 1 & 0 & ... & 0 & 0 \\
            0 &  0  & 0 & 0 & 1 & ... & 0 & 0 \\
            ... & ... & ... & ... & ... & ... & ... & ... \\
            0 &  0  & 0 & 0 & 0 & ... & 0 & 1 \\
            p &  1-p  & 0 & 0 & 0 & ... & 0 & 0 \\
          \end{bmatrix},
\end{align*}
\begin{align*}
    M_1 = \begin{bmatrix}
            0 &  0  & 1 & 0 & 0 & ... & 0 & 0\\
            p & 1-p & 0 & 0 & 0 & ... & 0 & 0\\
            0 &  0  & 0 & 1 & 0 & ... & 0 & 0 \\
            0 &  0  & 0 & 0 & 1 & ... & 0 & 0 \\
            ... & ... & ... & ... & ... & ... & ... & ... \\
            0 &  0  & 0 & 0 & 0 & ... & 0 & 1 \\
            p &  1-p  & 0 & 0 & 0 & ... & 0 & 0 \\
          \end{bmatrix}.
\end{align*}

Finally, our objective is to minimize the expected AoI for all users. Specifically, let AoI represent the steady-state AoI of a user as a random variable. We seek to optimize the parameters of the Markov transition matrices $M_0$ and $M_1$ by solving the following optimization problem:
\begin{align}
    \min \; & F(M_0, M_1) := E[AoI], \label{opt_problem}
\end{align}
where $E[\cdot]$ denotes the expectation function.

With the system's key parameters defined, we now turn to the next step of developing a solution framework that effectively models and minimizes AoI.


\section{Analysis through Second-Order Mean-Field Approximation}
\label{sec:solution}

Optimizing Age of Information (AoI) in random access networks presents significant challenges due to the complex interaction between users competing for transmission opportunities. To address this, we propose to combine second-order analysis with mean-field approximation. This method efficiently models and minimizes AoI while maintaining computational feasibility, ensuring both accuracy and practicality.

\subsection{Second-Order Approximation of AoI}

In the first step, we apply the second-order approximation method, as outlined in Guo \textit{et al.} \cite{guo2022theory} and Fan \textit{et al.} \cite{fan2023minimizing, fan2024optimizing}, to capture both the mean and temporal variance of packet deliveries. Since AoI is closely linked to the frequency and consistency of successful packet transmissions, this method approximates AoI by considering these two factors.

The packet delivery process for each user is represented by an indicator function $D_n(t)$, where $D_n(t) = 1$ if user $n$ successfully transmits at time $t$, and $D_n(t) = 0$ otherwise. In a symmetric network, all users exhibit similar statistical transmission patterns, enabling us to define two key quantities:
\begin{itemize}
    \item Mean success rate $m$: the long-term average rate at which a user successfully transmits a packet.
    \item Temporal variance $v^2$: the variability in packet deliveries over time.
\end{itemize}
These quantities are mathematically expressed as:
\begin{align*}
    m &:= \lim_{T \to \infty } \frac{\sum_{t=1}^{T}D_n(t)}{T}, \\
    v^2 &:= \lim_{T \to \infty} E\left[\left( \frac{\sum_{t=1}^{T}D_n(t)-T m}{\sqrt{T}}\right)^2\right].
\end{align*}

With these values, we approximate the expected AoI using the formula derived from Theorem 2 in \cite{fan2023minimizing}:
\begin{align}
    E[AoI] \approx \frac{1}{2}\left(\frac{v^2}{m^2} + \frac{1}{m}\right) + \frac{1}{2}. \label{eq:aoi_appr}
\end{align}

This approximation incorporates both the mean rate of successful transmissions and their variability, providing a more accurate estimate of the AoI. The remaining task is to compute $m$ and $v^2$ based on the transition matrices $M_0$ and $M_1$, which we achieve in the next step using mean-field approximation.

\subsection{Mean-Field Approximation of the Delivery Process}

In the second step, we focus on calculating the mean $m$ and variance $v^2$ by addressing the complexity of user interactions among users in large random access networks. As the number of users $N$ increases, directly modeling the behavior of each user becomes computationally intractable. 

To address this, we apply the mean-field approximation, which simplifies the analysis by replacing the interactions between individual users with the aggregate effect of the remaining $N - 1$ users. This approximation assumes that as the number of users increases, the impact of each user on the system diminishes.

Recall each user in the network follows a Markov process to transition between states (e.g., idle, transmission, or backoff), with transitions governed by the results of carrier sensing or acknowledgment (ACK) reception. The system's state is captured by the steady-state probability vector $\mu = \{\mu_1, \mu_2, \dots, \mu_S\}$, where $\mu_i$ represents the probability of a user being in state $i$ in the long run.

The system’s steady-state behavior is then described by the equation:
\begin{align}
\gamma_0 \mu M_0 + \gamma_1 \mu M_1 = \mu.
\label{eq:steady}
\end{align}
Here, $\gamma_0 = (1 - \mu_1)^{N-1}$ represents the probability that the channel is idle (i.e., no other users are transmitting), and $\gamma_1 = 1 - \gamma_0$ represents the probability that the channel is busy. The matrices $M_0$ and $M_1$ describe state transitions when the channel is idle or busy, respectively.

To solve for $\mu$, we employ a fixed-point iteration method. Starting with an initial guess for $\mu$, we iteratively update the value until it converges:
\begin{algorithm}
\caption{Fixed-Point Iteration for Steady-State Probability Calculation}
\begin{algorithmic}[1]
\renewcommand{\algorithmicrequire}{\textbf{Input:}}
\REQUIRE $M_0, M_1$
\STATE Initialize state distribution, e.g., $\Bar{\mu} = [1, 0, \dots, 0]$.
\STATE Compute $\gamma_0 = (1 - \Bar{\mu}_1)^{N-1}$.
\STATE Solve $\mu'$ from equation (\ref{eq:steady}) subject to $\sum_{s=1}^{S} \mu'_s = 1$.
\IF {$\|\Bar{\mu} - \mu'\| \geq$ threshold}
    \STATE Set $\Bar{\mu} = \mu'$ and return to step 2.
\ELSE
    \STATE Output $\mu = \mu'$.
\ENDIF
\renewcommand{\algorithmicrequire}{\textbf{Output:}}
\REQUIRE $\mu$
\end{algorithmic}
\label{alg:fixed_point}
\end{algorithm}

Once the steady-state probability vector $\mu$ is obtained, the mean success rate $m$ can be computed using the following expression:
\begin{align}
    m = \mu_1 \gamma_0 = \mu_1 (1 - \mu_1)^{N-1}. \label{eq:mean}
\end{align}
The temporal variance $v^2$ is calculated by central limit theory using the covariance of transmission successes across time steps:
\begin{align*}
    v^2 = m - m^2 + 2 \sum_{k=2}^{\infty} \text{Cov}(D_n(k), D_n(1)).
\end{align*}
The covariance term $\text{Cov}(D_n(k), D_n(1))$ measures how packet transmissions at different times are correlated. It is expressed as:
\begin{align*}
    & \text{Cov}(D_n(k),D_n(1)) \\
    &= E[D_n(1)D_n(k)]-E[D_n(1)]E[D_n(k)] \\
    &=\big(Prob(D_n(k)=1|D_n(1)=1)-E[D_n(k)]\big) E[D_n(1)] \\
    &=\big(O_S M_0 (\gamma_0 M_0 + \gamma_1 M_1)^{k-2}-m\big)m.
\end{align*}
Thus, the temporal variance incorporates both short-term and long-term dependencies in transmission success, captured by the covariance across multiple time steps:
\begin{align}
    v^2 = m - m^2 + 2\sum_{k=2}^{\infty}\big(O_S M_0 (\gamma_0 M_0 + \gamma_1 M_1)^{k-2}-m\big)m. \label{eq:var}
\end{align}

Substituting results from Algorithm.~\ref{alg:fixed_point}, equations (\ref{eq:mean}) and (\ref{eq:var}) into the second-order approximation equation (\ref{eq:aoi_appr}), we derive a highly efficient and accurate method to estimate and minimize the AoI, even in large-scale networks with complex interaction.

To validate the effectiveness of our proposed solution, we apply it to the age-threshold ALOHA policy. The following section presents simulation results and compares the performance of our framework against existing parameter settings.


\section{AoI Optimization for Age-Threshold ALOHA Strategy}
\label{sec:results}

In this section, we optimize the age-threshold ALOHA policy parameters using our proposed algorithm and benchmark its performance against that of established parameter settings from the literature \cite{yavascan2021analysis}. The two key parameters in the age-threshold ALOHA policy are the age threshold $H$ and the ALOHA transmission probability $p$.

We evaluate three parameter settings for minimizing Age of Information (AoI). First, our proposed \textbf{Second-Order and Mean-field Analysis (SOMA)} approach combines second-order analysis with mean-field approximation, as detailed in previous sections, to estimate AoI across a range of values for $H$ (from 1 to $3N$) and $p$ (with a precision of 0.001). In practice, using $k$ values in (\ref{eq:var}) between 2 and 1000 provides sufficient accuracy, with a convergence threshold of $1 \times 10^{-6}$. This method selects the parameters that obtain minimum calculated AoI.

The second and third approaches are both derived from the reference paper \cite{yavascan2021analysis}. \textbf{Literature-Based Optimal Parameters (LBOP)} is said to be optimal in \cite{yavascan2021analysis}, with recommended values of $H = 2.2N$ and $p = 4.69/N$. LBOP assumes a two-peak behavior, where the system oscillates between two distinct peaks of active users, balancing collision probability and channel utilization to achieve optimal performance in larger networks. In contrast, \textbf{Single-Peak Guided Parameters (SPGP)}, with $H = 2.17N$ and $p = 4.43/N$, assumes a single-peak behavior, stabilizing the number of active users around a single value. This setting is better suited for smaller networks, where reduced collisions and improved stability lead to better AoI performance

\begin{figure}[htbp]
\centerline{\includegraphics[scale=0.48]{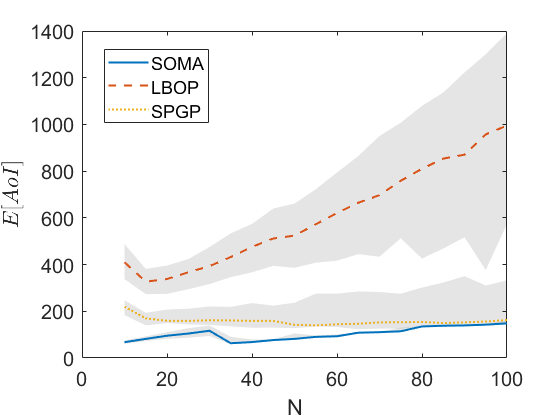}}
\caption{Expected AoI Comparison Among SOMA, LBOP, and SPGP.}
\label{fig:aoi_comp}
\end{figure}

We evaluate the practical performance of these parameter settings by simulating the system for 100,000 time slots across 100 independent runs. The key metric is the expected AoI, averaged over all runs, and the 95\% percentile interval is shown as a gray area, as illustrated in Figure~\ref{fig:aoi_comp}.

\vspace{-10pt}
\vspace{-5pt}

As shown in Figure~\ref{fig:aoi_comp}, our proposed optimal parameters derived from SOMA consistently outperform LBOP and SPGP across varying network sizes $N$. SOMA achieves a significantly lower expected AoI, demonstrating the effectiveness of our approach in practical scenarios. Additionally, the 95\% percentile intervals show that SOMA offers more stable performance, while both LBOP and SPGP exhibit higher variability, further emphasizing the robustness of SOMA.

A very surprising result in Figure~\ref{fig:aoi_comp} is that LBOP and SPGP perform worse than our setting based on SOMA, despite that LBOP and SPGP have been shown to be the asymptotic optimal setting under some assumptions \cite{yavascan2021analysis}. To gain better understanding why they perform poorly, we evaluate the performance of the age-threshold ALOHA policy under different $\epsilon:=p/N$ while setting $H$ to be $2.2N$. Note that LBOP suggests choosing $\epsilon=4.69$ and SPGP suggests choosing $\epsilon=4.43$.

\begin{figure}[htbp]
\centerline{\includegraphics[scale=0.48]{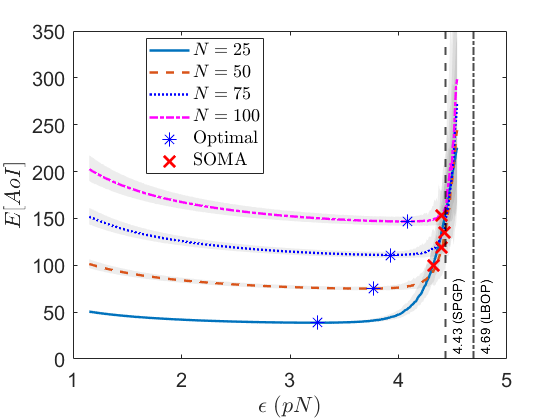}}
\caption{Expected AoI Comparison for Different $\epsilon$.}
\label{fig:aoi_eps}
\end{figure}

Simulation results for different $H$ are shown in Figure~\ref{fig:aoi_eps}. It can be observed that the optimal $\epsilon$, that is, the $\epsilon$ with the smallest AoI, indeed converge to somewhere between 4.43 and 4.69. This shows that the analysis in \cite{yavascan2021analysis} is indeed valid. However, we also notice that the average AoI starts to increase sharply once $\epsilon$ becomes just a little larger than the optimal value. As a result, even though the choices of $\epsilon$ under LBOP and SPGP are only a little larger than the optimal values, they can still result in poor average AoI. In contrast, our setting based on SOMA is much more robust across different $N$.

\vspace{-10pt}
\vspace{-5pt}

Finally, we compare the computational runtime of our proposed algorithm with that of a practical simulation approach, as shown in Table~\ref{tb:runtime}. The runtime for practical simulations is the average of 80 trials, where each trial consists of 100,000 time slots and 100 independent runs.

\vspace{-10pt}
\begin{table}[htbp]
\caption{Runtime Comparison}
\begin{center}
\begin{tabular}{|c|c|c|c|c|c|}
\hline
$N$ & 25 & 50 & 75 & 100 \\
\hline
SOMA & 0.0025 & 0.0091 & 0.0187 & 0.0298 \\
\hline
Practical Simulation & 12.1159 & 22.3107 & 33.3154 & 44.6191 \\
\hline
\end{tabular}
\label{tb:runtime}
\end{center}
\end{table}
\vspace{-10pt}

As shown in the table, our proposed method is thousands of times faster than running full practical simulations, highlighting the efficiency of applying our solution for AoI optimization in real-time applications.

These findings highlight the practical significance of the proposed framework, which consistently outperforms existing approaches. The following section summarizes the key contributions of this approach.


\section{Conclusion}
\label{sec:conclusion}

In this paper, we introduce a novel framework to analyze and optimize Age-of-Information (AoI) in distributed random access networks utilizing CSMA protocols. By integrating second-order analysis with mean-field approximation, we develop a tractable approach that approximates AoI while accounting for complex user interference and supporting a wider range of transmission strategies, including arbitrary Markov processes. Through extensive simulations of the recently proposed age-threshold policy, we demonstrate that our approach’s derived parameters outperform those from the asymptotic analysis of previous works, highlighting the adaptability and effectiveness of the proposed framework. This makes it a valuable tool for both theoretical analysis and practical implementations in time-sensitive communication systems.


\bibliographystyle{ieeetr}
\bibliography{Main}

\end{document}